# Tailoring electronic and elastic properties by varying composition of the $CuGa_{1-x}Al_xS_2$ chalcopyrite semiconductor

**M.G. Brik*, C.-G. Ma**

Institute of Physics, University of Tartu, Riia 142, Tartu 51014, Estonia

## Abstract

Influence of composition and external hydrostatic pressure on the structural, electronic, and optical properties of the $CuGa_{1-x}Al_xS_2$ ($x$=0, 0.25, 0.5, 0.75, 1.0) chalcopyrite semiconductor was analyzed by means of the first-principles calculations. Dielectric functions and optical absorption spectra were calculated for all considered aluminum concentrations. The pressure coefficients of the calculated band gaps and position of the lowest in energy absorption peaks were extracted from the calculated results. One of the main results is that substitution of 25% of gallium by aluminum (thus forming the $CuGa_{0.75}Al_{0.25}S_2$ semiconductor) increases absorption in the visible part of the solar spectrum by about 6%, which can be important for the solar cell applications.

## 1. Introduction

A search for renewable ecological resources of energy is one of the greatest challenges for a modern industrial society, heavily dependent on production and consumption of electrical energy. One of possible alternatives to the traditional thermal power stations, for example, is a solar cell, converting the energy of Sun into electricity. Although silicon solar panels are widespread, searches for new materials and attempts of improving their performance and efficiency have never stopped. It is a common understanding now that the materials for the potential solar cell applications should be semiconductors with rather narrow band gap matching or close to the visible part of solar spectrum. Among various compounds, already used and still tested for the photovoltaic applications, the I-III-$VII_2$ ternary semiconductors with the chalkopyrite structure (e.g. $CuGaS_2$, $CuInS_2$ etc) are considered as very suitable materials, since they can be grown in the form of thin films [1, 2, 3, 4, 5] to increase the surface exposed to the sunlight. There is quite a number of theoretical works reporting results of the first-principles calculations for $CuGaS_2$ [6, 7, 8, 9, 10, 11, 12, 13 etc], clarifying details of the electronic structure and optical properties of this representative of a large family of ternary semiconductors with chalkopyrite structure. The experimental band gap value of $CuGaS_2$ is about 2.43 eV [14], and in attempt to increase efficiency of the solar light absorption doping with various elements (C, Si, Ge, Sn or d-metals) has been suggested [15, 16]. Such doping would lead to creation of the interband states, which would allow absorbing the low energy photons from the Sun spectrum.

* Corresponding author. E-mail: brik@fi.tartu.ee

So far, not too many works focused on the variation of the second cation in the chalcopyrite semiconductors were published. Thus, the only known to us experimental work on the $CuAl_xGa_{1-x}S_2$ alloys reported the crystal growth of the above-mentioned films [17]. We also mention the experimental studies of the spectroscopic properties of $CuGa_xIn_{1-x}S_2$ crystals [18].

In the present work we continue our previous studies of the neat I-III-$VII_2$ materials [10] and consider how the substitution of the Ga atoms by Al ions would modify the structural, electronic, optical and elastic properties of the $CuGa_{1-x}Al_xS_2$ ($x$=0...1) compound, leading to overall enhancement of the absorptive abilities of a new mixed material. A partial or a complete substitution of Ga by Al ions does not lead to any structural modification other than scaling of the lattice parameters and requires no charge compensating defects, since both ions are in the same oxidation state +3, and both "pure" $CuAlS_2$ and $CuGaS_2$ compounds have the same crystal structure. In addition, we study the effect of the external hydrostatic pressure on absorption in the visible and ultraviolet spectral ranges. We have established how the lattice constants of these "mixed" materials and their band gaps depend on pressure. The main result is that substitution of 25 % of Ga ions in $CuGaS_2$ by Al ions would enhance absorption properties of $CuGa_{0.75}Al_{0.25}S_2$ by about 6%, when compared to the Al-free $CuGaS_2$; it is worthwhile noting that the position of the maximum of the visible absorption band in that mixed $CuGa_{0.75}Al_{0.25}S_2$ compound is still kept very close to the maximum of solar spectrum.

In the next section we describe the structure of the considered semiconductors and the details of calculations, then we proceed with the description of the calculated results and we finish the paper with giving a short summary.

## 2. Crystal structure and details of calculations

One unit cell of $CuGaS_2$ (isostructural to $CuAlS_2$) is shown in Fig. 1. This is a typical example of the chalkopyrite structure, space group I-42d with four formula units per one unit cell; each atom has four nearest neighbors: every metal ion is coordinated by four sulfur ions, every sulfur ion has two Ga and two Cu nearest neighbors. The crystal structure data are collected in Table 1.

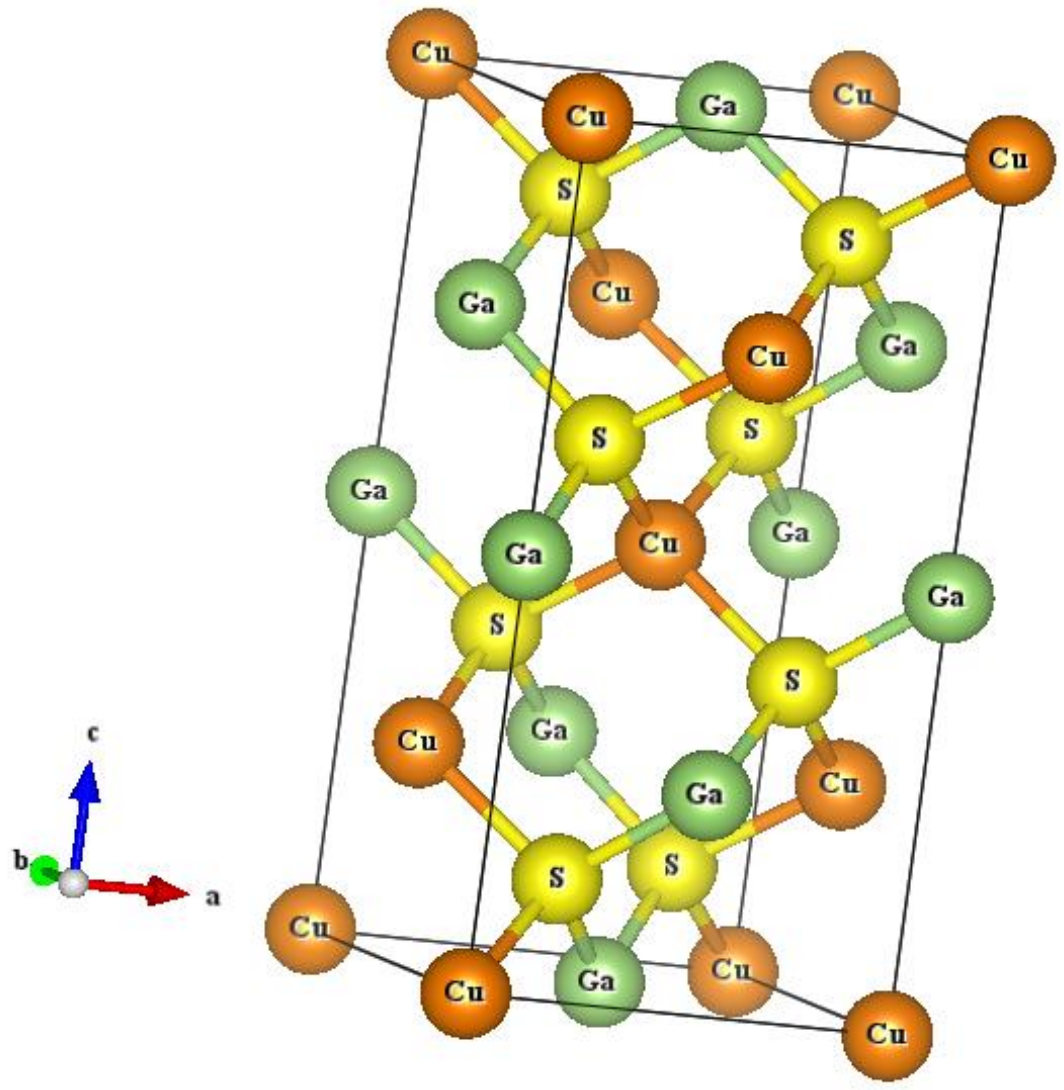

Fig. 1. One unit cell of $CuGaS_2$. Drawn with VESTA [19].

All calculations were performed using the CASTEP module [20] of the Materials Studio package. The exchange–correlation effects were treated within the generalized gradient approximation (GGA) with the Perdew–Burke–Ernzerhof functional [21]. The plane wave basis set cut-off energy was chosen at 350 eV, the Monkhorst–Pack scheme *k*-point grid sampling was set as 5×5×3 k-points for the Brillouin zone. The convergence tolerance parameters were: energy $10^{-5}$ eV/atom, maximal force and stress 0.03 eV/Å and 0.05 GPa, respectively, and the maximal displacement 0.001 Å. The calculations were performed for a conventional cell, in which four Al atoms were gradually replaced (one by one) by Ga atoms. The considered electronic configurations were $3d^{10}4s^1$ for Cu, $3d^{10}4s^24p^1$ for Ga, $3s^23p^4$ for S, and $3s^23p^1$ for Al.

We have also studied the pressure effects on the structural, optical and electronic properties of the whole series of the $CuGa_{1-x}Al_xS_2$ crystals for $x$=0, 0.25; 0.5; 0.75, 1; to do that, all calculations were repeated for all compounds in the pressure range from 0 to 20 GPa with a step of 5 GPa. The obtained results are described and discussed in the next sections.

## 3. Results of calculations

### 3.1. Structural properties

Table 1 collects the calculated (in this and other available works) lattice parameters of $CuGa_{1-x}Al_xS_2$ crystals in comparison with the experimental findings. As seem from the Table, agreement between the present and experimental results is very good, with the maximum relative error of about 1.2 % for the *a* parameter of $CuAlS_2$ and less than 1 % in all remaining cases, thus indicating high accuracy of the performed calculations and reliability of the results obtained for these optimized crystal structures.

Since no literature data were found for the mixed compounds with the varying content of both Ga and Al, the obtained parameters can not be compared with any other results. However, the overall consistency of the data presented in Table 1 is confirmed by Fig. 2, which shows a linear dependence of the lattice parameters *a* and *c* on the Al content *x* in $CuGaS_2$.

Table 1. Summary of structural properties for $CuGa_{1-x}Al_xS_2$ crystals

| | $x$=0.0 | $X$=0.25 | $x$=0.5 | $x$=0.75 | $x$=1.0 |
|---|---|---|---|---|---|
| $a$, Å | 5.3312[a], 5.351[b], 5.356[c], 5.3819[d] 5.3512[e] | 5.3161[a] | 5.3018[a] | 5.2836[a] | 5.2689[a], 5.3336[b] 5.341[f,h], 5.321[g] 5.2389[i] |
| $c$, Å | 10.5794[a], 10.480[b] 10.629[c], 10.660[d] 10.478[e] | 10.5328[a] | 10.4855[a] | 10.4569[a] | 10.4179[a], 10.444[b] 10.513[d], 10.525[g] 10.570[h], 10.4148[i] |

[a] Calc., present work.
[b] Experiment, Ref. [22]
[c] Calc., Ref. [10]
[d] Calc., Ref. [12]
[e] Calc., Ref. [13]
[f] Calc., Ref. [23]
[g] Calc., Ref. [24]
[h] Calc., Ref. [25]
[i] Calc., Ref. [26]

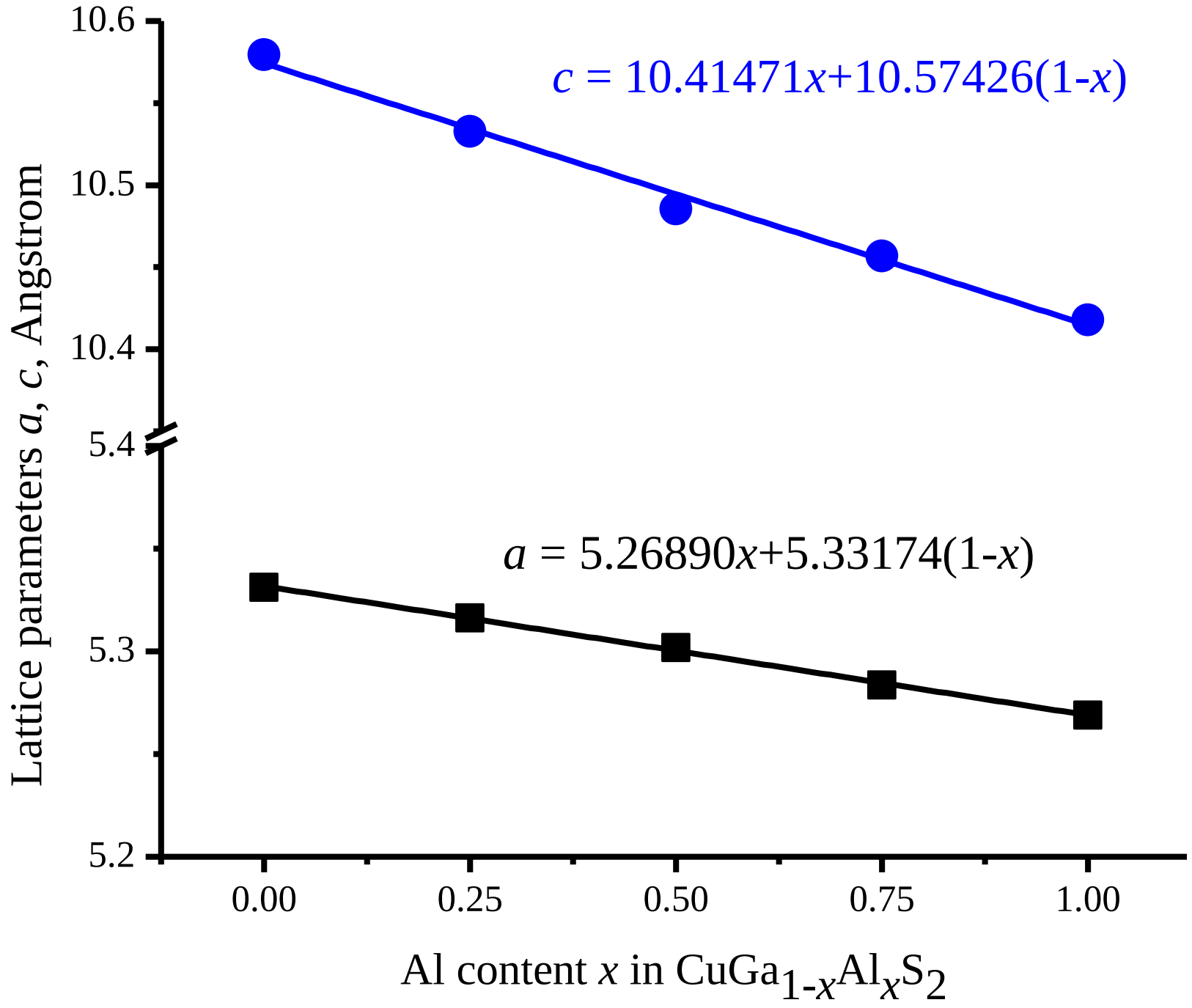

Fig. 2. Dependence of the calculated lattice parameters (symbols) on the Al content in $CuGa_{1-x}Al_xS_2$. Equations of the fitting straight lines are shown.

As Fig. 2 shows, there are perfectly linear dependencies of both lattice parameters on the Al content, as can be anticipated from the Vegard's law [27]. It is easy to see that the two fitting coefficients of each composition dependencies in Fig. 2 correspond to the lattice parameters of pure $CuAlS_2$ and $CuGaS_2$ crystals, respectively. Gradual substitution of the Ga ions by the Al ions is accompanied by a decrease of the lattice parameters, since the ionic radius of $Al^{3+}$ in the four-fold coordination (0.39 Å) is smaller than the radius of $Ga^{3+}$ (0.47 Å) [28].

3.2. Electronic properties

The crystals considered in the present work are all the direct band gap materials; the maximum of the valence band and the minimum of the conduction band are realized at the Γ point (center of the Brillouin zone). The calculated band gaps for pure $CuGaS_2$ and $CuAlS_2$ turned out to be 0.8 eV and 1.96 eV, respectively. Both values are underestimated if compared to the experimental results of 2.43 eV [14] and 3.50 eV [29]. This underestimation is a well-known feature of the GGA approach, and can be overcome by introducing a scissor operator [30], which simply shifts upward the conduction band. In our case, the value of such a shift was 1.55 eV for both crystals.

Table 2. Summary of electronic and elastic properties for $CuGa_{1-x}Al_xS_2$ crystals

| | $x$=0.0 | $x$=0.25 | $x$=0.5 | $x$=0.75 | $x$=1.0 |
|---|---|---|---|---|---|
| Band gap, eV | 0.8[a], 0.903[b], 0.818[c], 1.1[d] | 1.03[a] | 1.31[a] | 1.45[a] | 1.96[a], 2.05[e], 1.718[f], 2.29[g] |
| Bulk modulus, GPa | 78.4[a], 96.2[h], 88.9[i], 75.1[j] | 80.1[a] | 82.2[a] | 83.9[a] | 85.3[a], 84.1[k] |
| $B' = dB/dP$ | 4.4[a], 4.8[b] | 4.4[a] | 4.3[a] | 4.3[a] | 4.2[a], 4.6[k] |

[a] Present work.
[b] Calc., Ref. [9]
[c] Calc., Ref. [31]
[d] Calc., Ref. [13]
[e] Calc., Ref. [32]
[f] Calc., Ref. [24]
[g] Calc., Ref. [33]
[h] Calc., Ref. [9]
[i] Exp., Ref. [34]; [j] Calc., Ref. [35]; [k] Calc., Ref. [25]

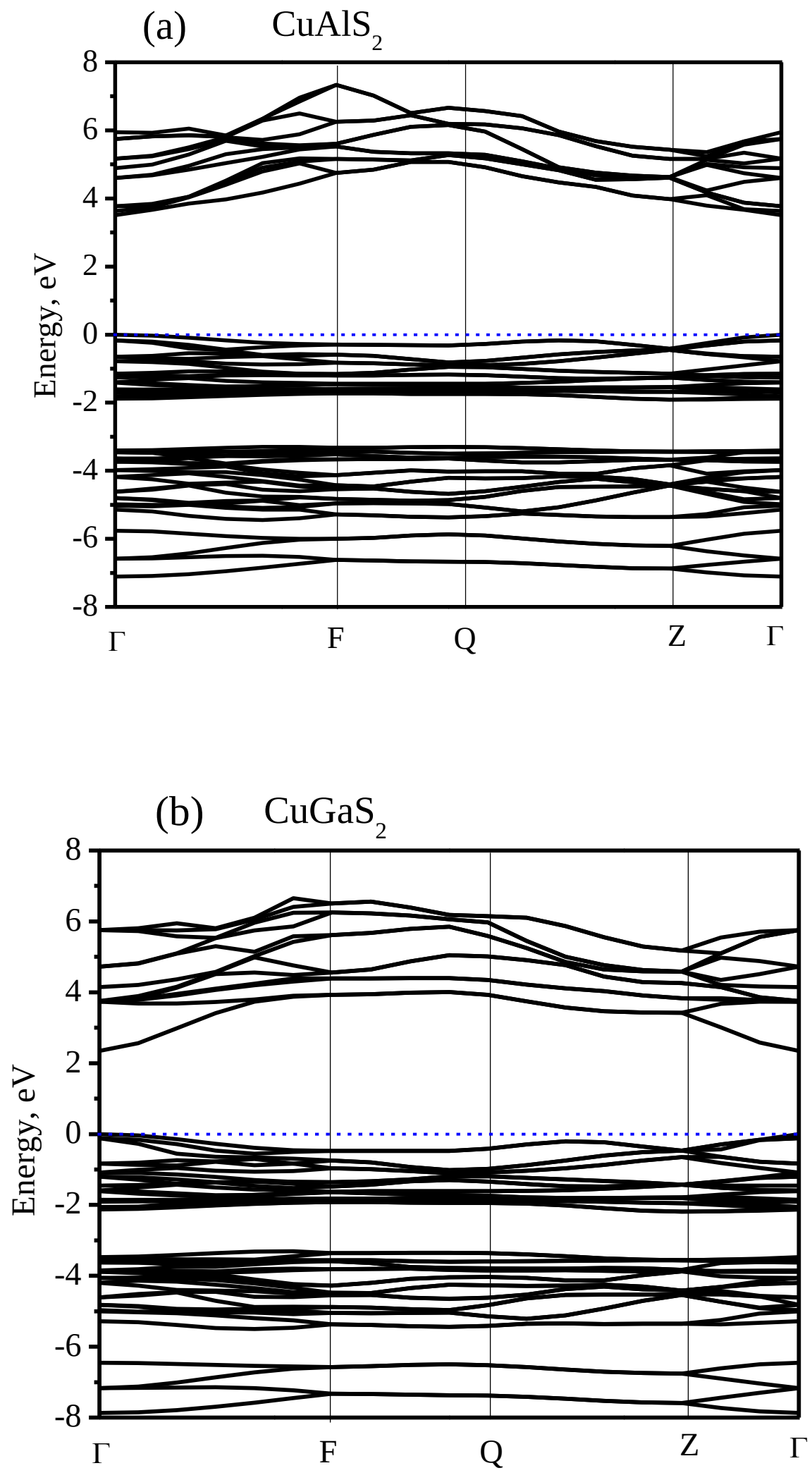

Fig. 3. Calculated band structures for pure $CuAlS_2$ (a) and $CuGaS_2$ (b) (1.55 eV scissor operator is applied). The coordinates of the special points of the Brillouin zone are (in unit vectors of the reciprocal lattice): $\Gamma$(0, 0, 0); F (0, 1/2, 0); Q (0, 1/2, 1/2), Z (0, 0, 1/2).

The conduction bands exhibit a well-pronounced dispersion and have similar widths of about 4.5 eV. The upper valence bands of both pure crystals are very flat; they have a very close width of about 2 eV; the lower valence band stretches from -3 eV to -7-8 eV. Such a structure is caused by splitting of the Cu 3d states, as was discussed in Ref. [10].

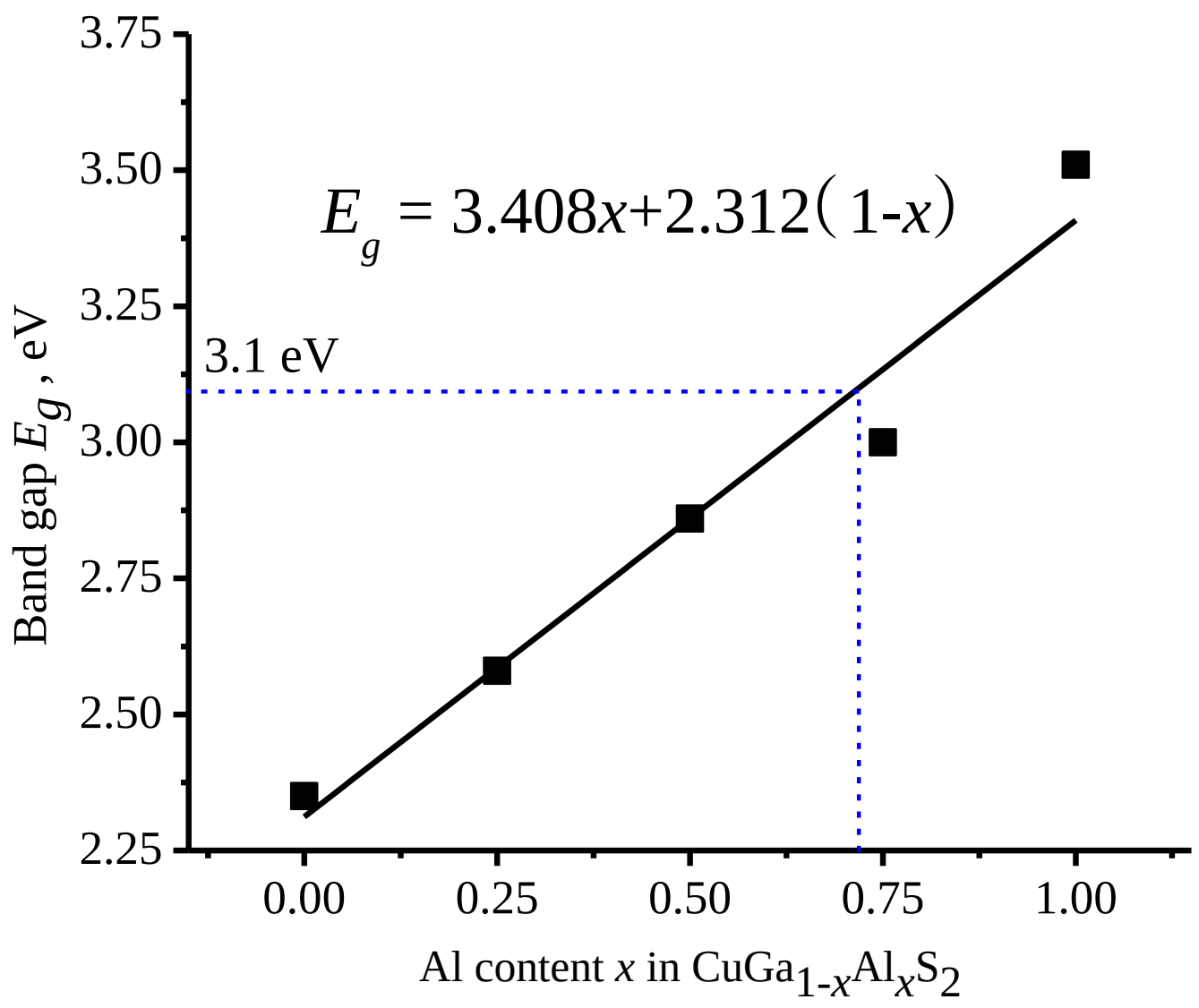

Fig. 4. Dependence of the calculated band gap (with 1.55 eV scissor) on the Al content in $CuGa_{1-x}Al_xS_2$. The aluminum threshold concentration corresponding to the visible light absorption (3.1 eV) is indicated by a vertical line.

Increasing amount of Al in $CuGa_{1-x}Al_xS_2$ leads to widening the band gap, as is demonstrated by Fig. 4. The dependence is fairly linear, with the fitting equation shown in Fig. 4. The fitting coefficients in comparison with the band gap values of pure $CuAlS_2$ and $CuGaS_2$ crystals suggest that the Vegard's law is still valid for the composition dependence of the electronic properties. Starting from pure $CuAlS_2$ with its UV band gap (i.e., the case for $x$=1 in Fig. 4), by adding Ga instead of Al it is possible to match the band gap of such a mixed material with the "blue" limit of visible spectrum at about 3.1 eV; such a condition can be realized for the $CuGa_{0.28}Al_{0.72}S_2$ compound as was estimated in Fig. 4. Therefore, this material and all other mixtures with increasing amount of Ga can be in principle used for solar cell applications, since they effectively absorb visible light.

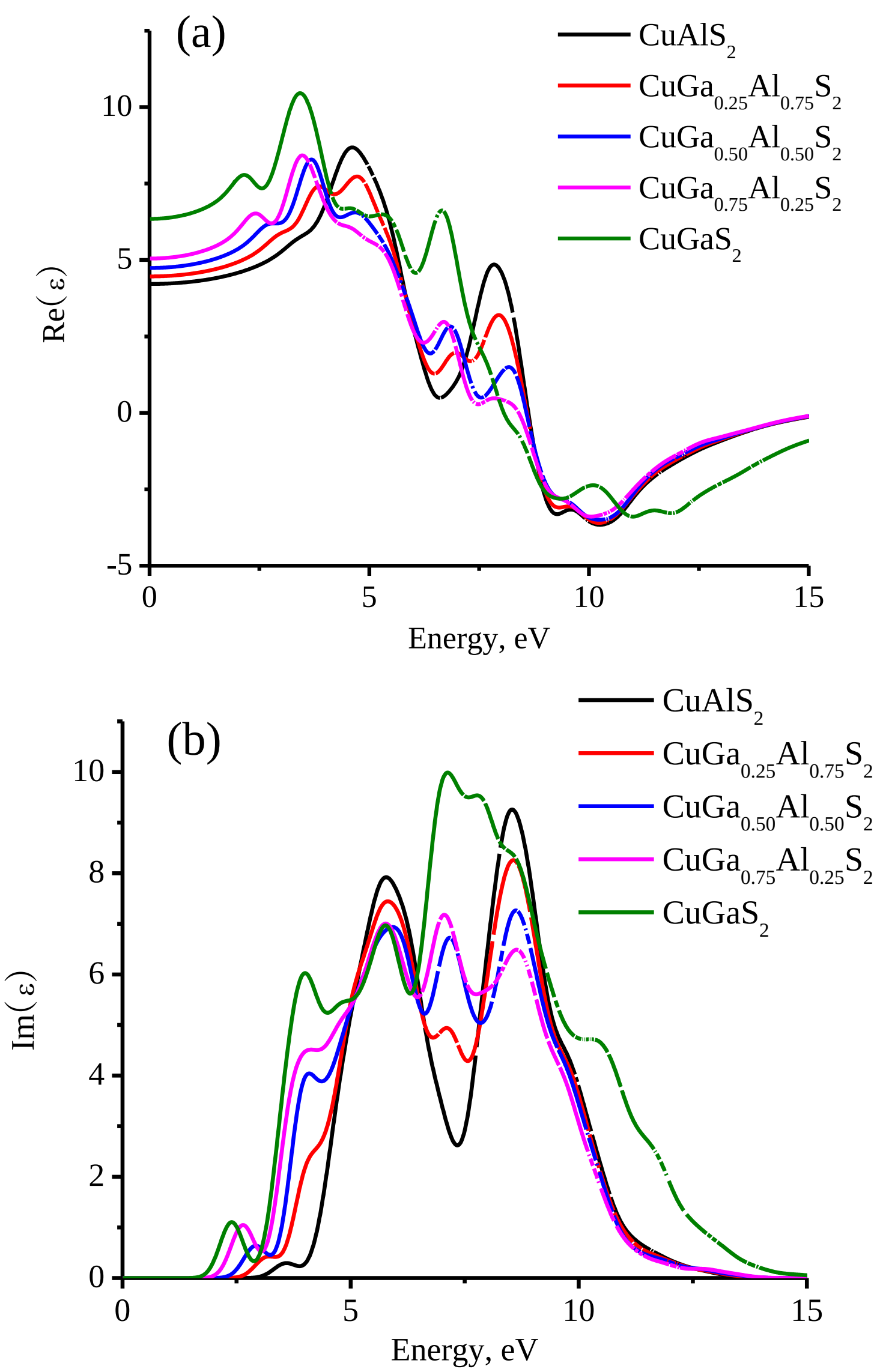

Fig. 5. Calculated real (a) and imaginary (b) parts of the dielectric function (with 1.55 eV scissor) for $CuGa_{1-x}Al_xS_2$ crystals.

Systematic variation of the real and imaginary parts of dielectric function for the $CuGa_{1-x}Al_xS_2$ crystals is given in Fig. 5. It can be noted that the Re($\varepsilon$) at zeroth energy decreases when moving from pure $CuGaS_2$ to pure $CuAlS_2$; the same trend was reported in Ref. [10] for the refractive indexes. At the same time, variation of the imaginary part of dielectric function reflect the changes of the band gap with increase of the Al concentration: the rising non-zero part of Im($\varepsilon$) is shifting to the higher energies.

The calculated absorption spectra in the visible range for all materials in the considered series are depicted in Fig. 6.

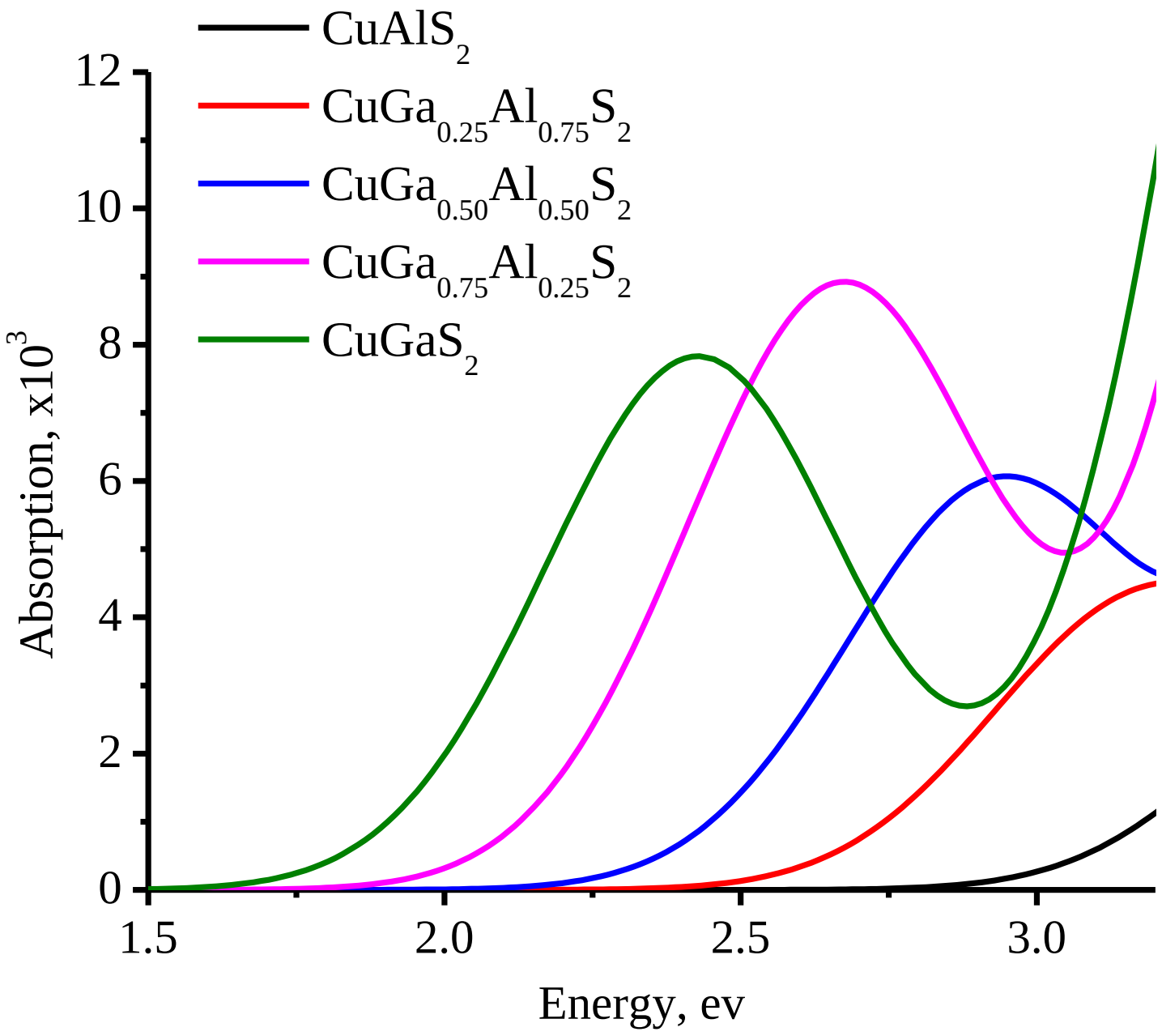

Fig. 6. Calculated absorption spectra in the visible range (with 1.55 eV scissor) for $CuGa_{1-x}Al_xS_2$ crystals.

An interesting feature of the shown in Fig. 6 spectra is that the visible absorption peak of $CuGa_{0.75}Al_{0.25}S_2$ has the maximum intensity among all considered five compounds. Increase of the absorption intensity, which was estimated by calculating the areas under absorption peaks, is about 6 %. At the same time, the blue shift of the visible absorption maximum is not large, still keeping it close to the maximum of the sunlight spectrum. Using the data presented in Figs. 4 and 6, it is also possible to suggest that a smaller Al concentration (about 12 %) would lead to the band gap of about 2.5 eV (~500 nm), which matches better the maximum of the Sun spectra than pure $CuGaS_2$, with simultaneous enhancement of the host absorption.

### 3.3. Pressure effects on the structural, electronic and optical properties

Optimization of the crystal lattice structures of all five compounds was performed in the pressure range from 0 to 20 GPa with a step of 5 GPa. Variations of the relative change of a unit cell volume $V/V_0$ (with $V$ and $V_0$ being the unit cell volume at pressure $P$ and ambient pressure) were fitted then to the Murnaghan equation of state [36]

$$\frac{V}{V_0} = \left(1 + P\frac{B'}{B}\right)^{-\frac{1}{B'}} \tag{1}$$

where $B$ and $B' = dB/dP$ are the bulk modulus and its pressure derivative, respectively, giving a simple way of extracting the elastic parameters from the fit.

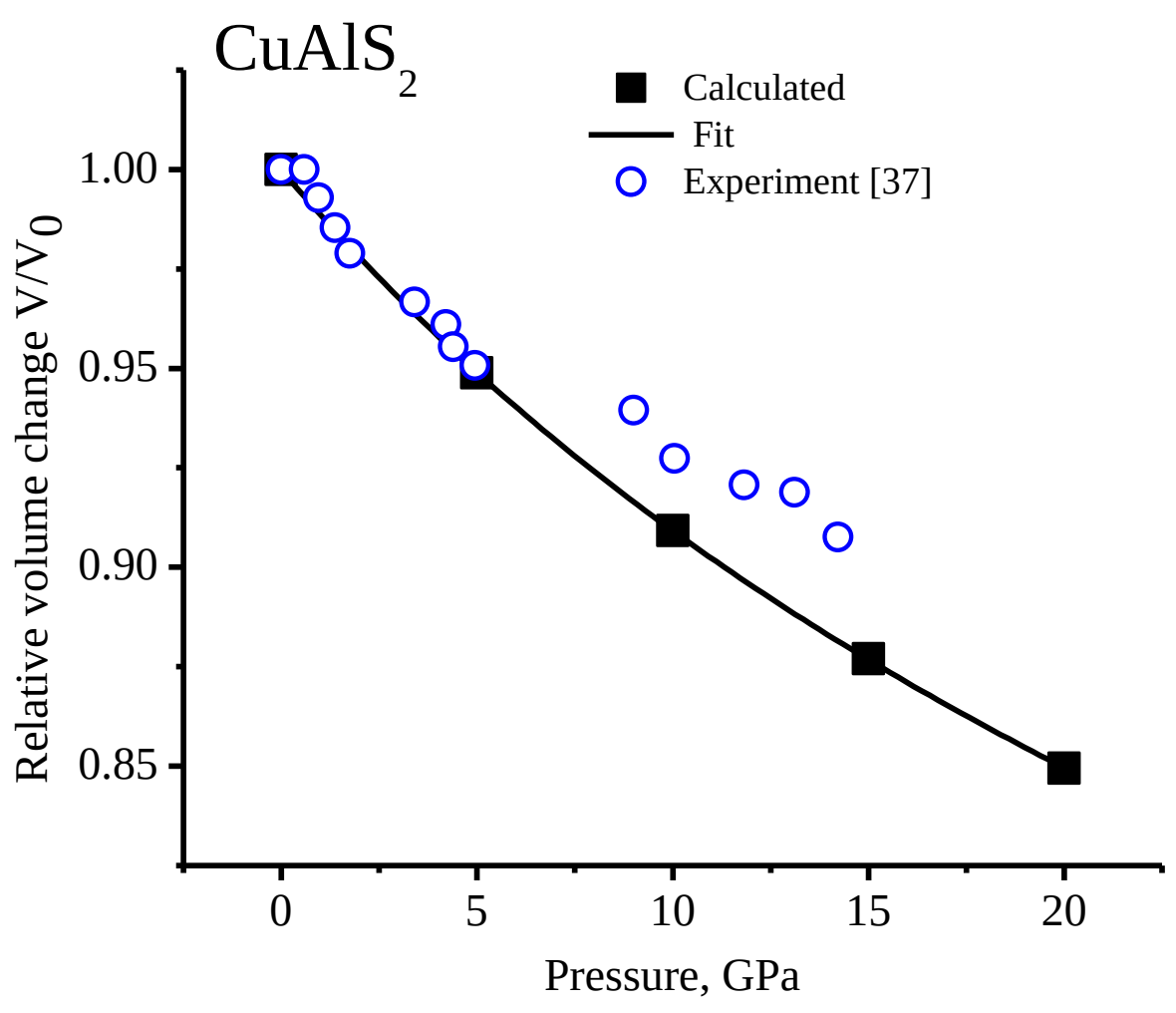

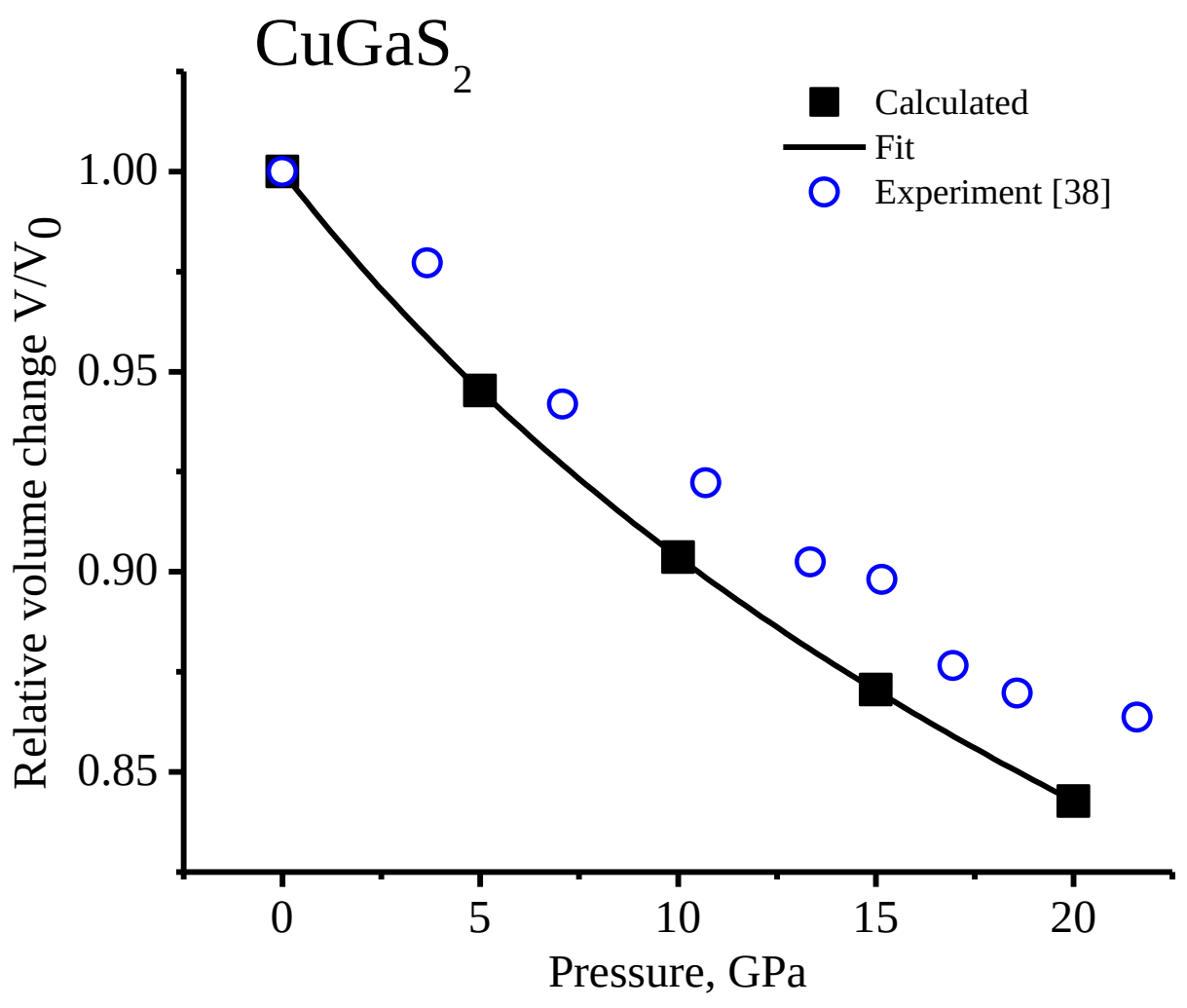

Fig. 7. Calculated variations of the relative volume change (filled squares), fits to the Murnaghan equation of state (lines) and experimental data (open circles) for $CuAlS_2$ and $CuGaS_2$.

The results of these calculations for the neat $CuAlS_2$ and $CuGaS_2$ are presented in Fig. 7, in comparison with the experimental data from Refs. [37, 38], respectively. Agreement with the experimental data for $CuAlS_2$ is very good until 5 GPa, and became somewhat worse for higher pressures. For $CuGaS_2$ agreement with the experimental data is moderate, but we stress out a rather wide range of the reported values of the bulk modulus *B* for this material (Table 2), which can be a source of deviation between the theory and experiment. The calculated values of the bulk moduli obtained in the present work for neat $CuAlS_2$ and $CuGaS_2$ are all consistent with other literature data.

Smooth variation of chemical composition in the $CuGa_{1-x}Al_xS_2$ series is followed by a continuous change of the bulk modulus *B*, as is shown in Fig. 8. The value of *B* is a linear function of the aluminum content *x*, which is completely similar with the cases of the structural and electronic properties. Dependence of the bulk modulus on the inverse volume of a unit cell for the whole series of the studies compounds is also given in Fig. 8. It is a linear function as well; a similar result was obtained in Ref. [39] for TlN, TlP, TlAs. The linear equations shown in Fig. 8 allow to determine the value of the bulk modulus for any concentration of Al in $CuGa_{1-x}Al_xS_2$ in the range $0 \le x \le 1$ or any unit cell volume.

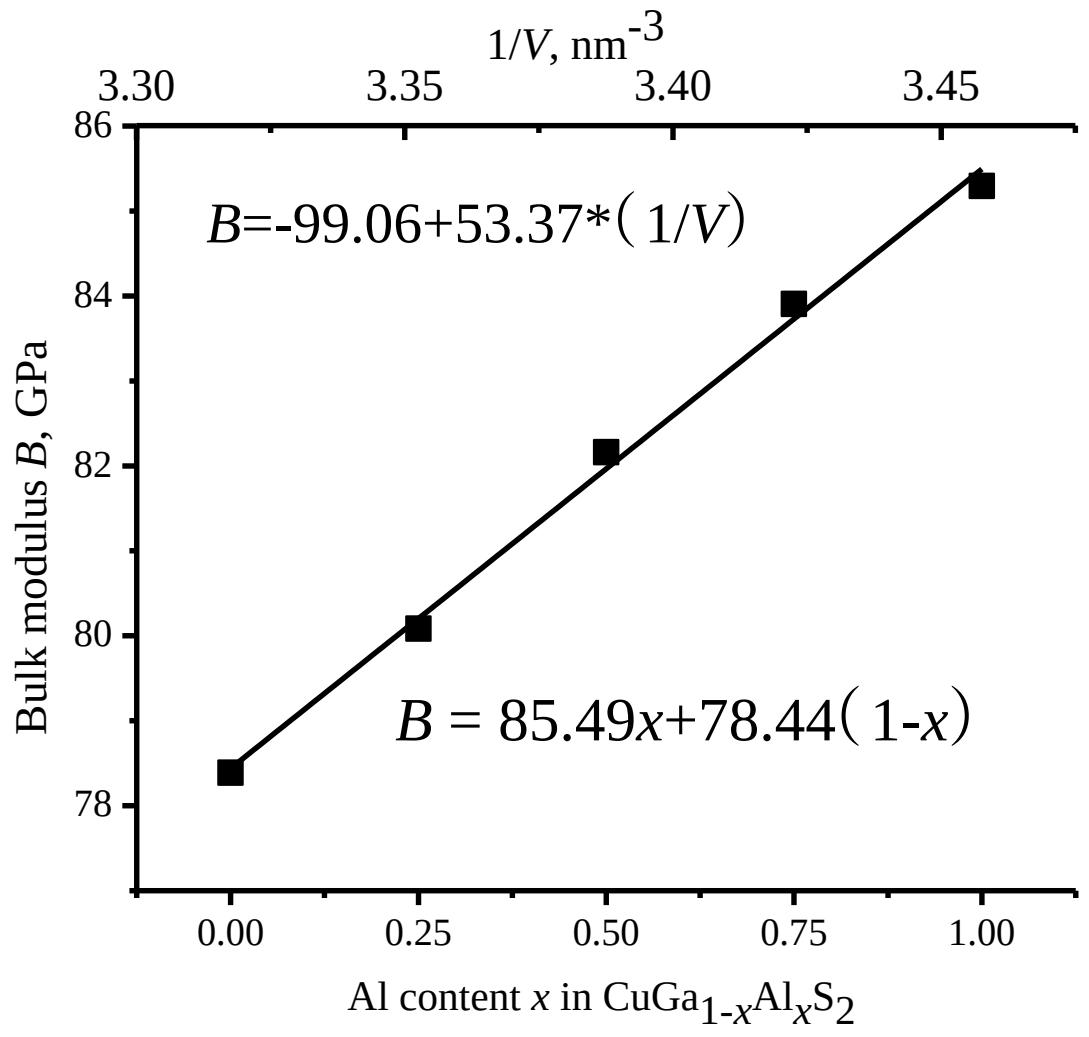

Fig. 8. Dependence of the bulk modulus *B* on Al content and on the inverse volume of a unit cell for $CuGa_{1-x}Al_xS_2$ crystals.

It is well known that the external hydrostatic pressure applied to a direct band gap material would lead to the increased band gap value. Besides, since the crystal lattice ions under pressure are brought closer to each other in the sample, the electronic density distribution is modified and, as a result, the optical absorption spectra would be also affected as well. Fig. 9 shows the visible and near UV parts of the calculated absorption spectra for $CuGa_{1-x}Al_xS_2$ with $x$=0, 0.25, 0.5, 0.75, 1.0 at different hydrostatic pressures.

For each crystal considered there is a blue shift of the absorption, due to the increased direct band gap. Moreover, integral intensity of absorption is increased, on account of enhanced absorption at about 6-7 eV. It is interesting to follow how the first low-energy absorption band depends on pressure. Comparing the samples with different amount of Al at the same pressure, we see that increasing aluminum concentration shifts the first absorption peak toward higher energies. For each concentration of aluminum an increased pressure also shifts the first absorption maximum in the same direction. It is interesting that the position of the first absorption peak maximum depends linearly on the pressure; this is proved by Fig. 10 (for pure $CuAlS_2$ the first absorption peak is merged with the UV absorption at high pressure, that is why only four sets of data for $x$=0, 0.25, 0.5 and 0.75 are shown).

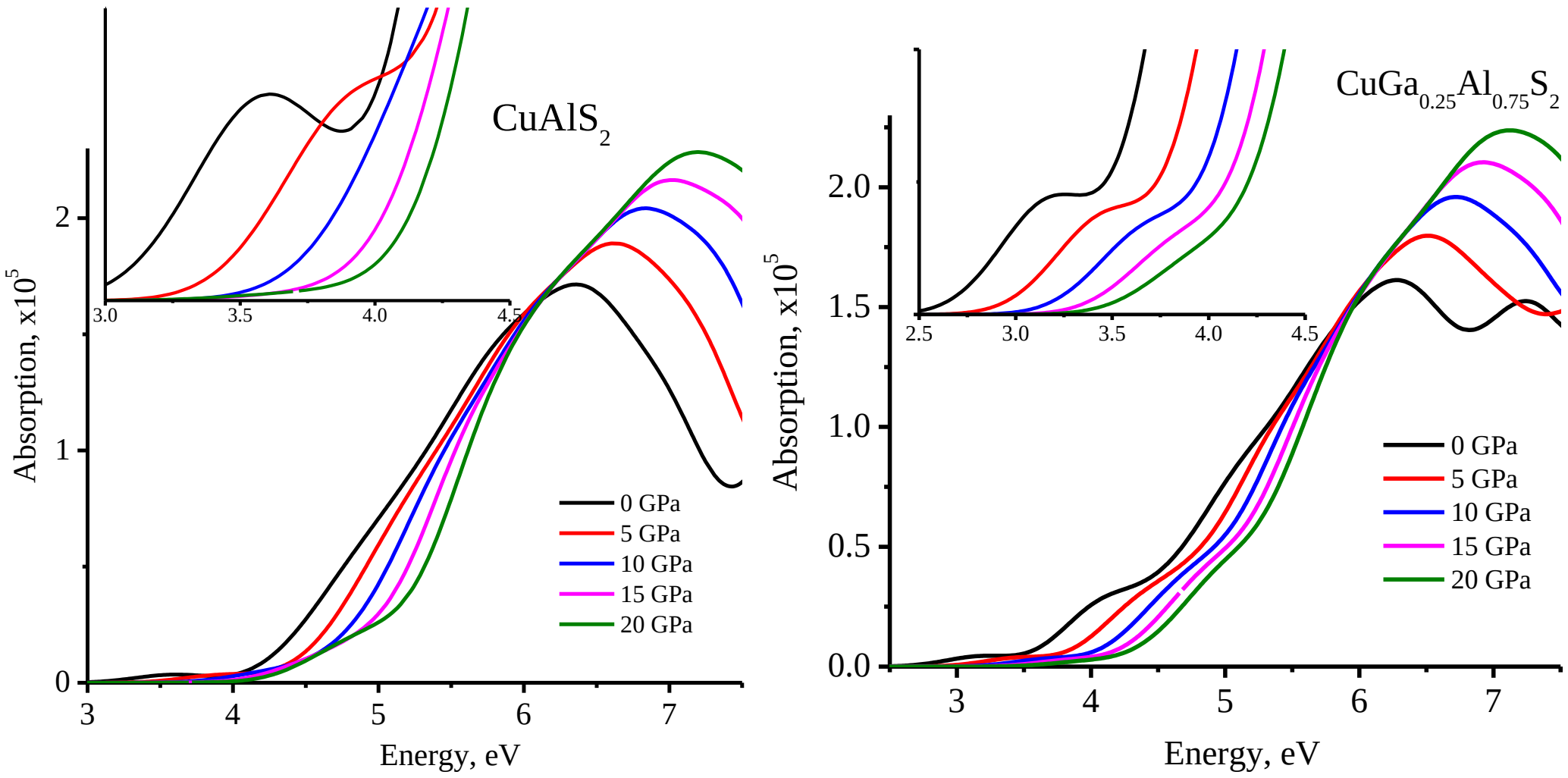

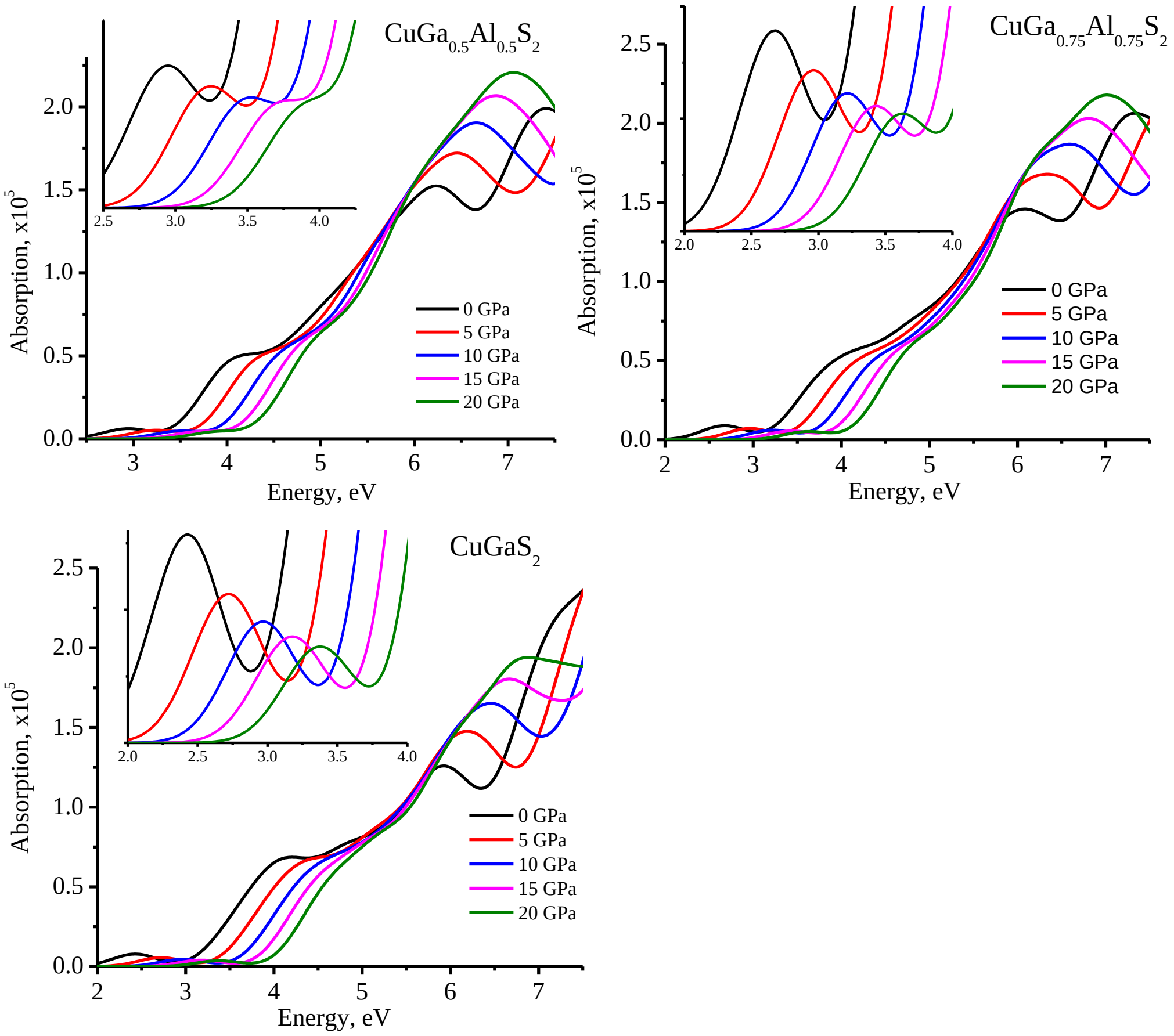

Fig. 9. Calculated visible and UV absorption spectra of $CuGa_{1-x}Al_xS_2$ crystals. Insets show the enlarged part in the range of the first absorption peak.

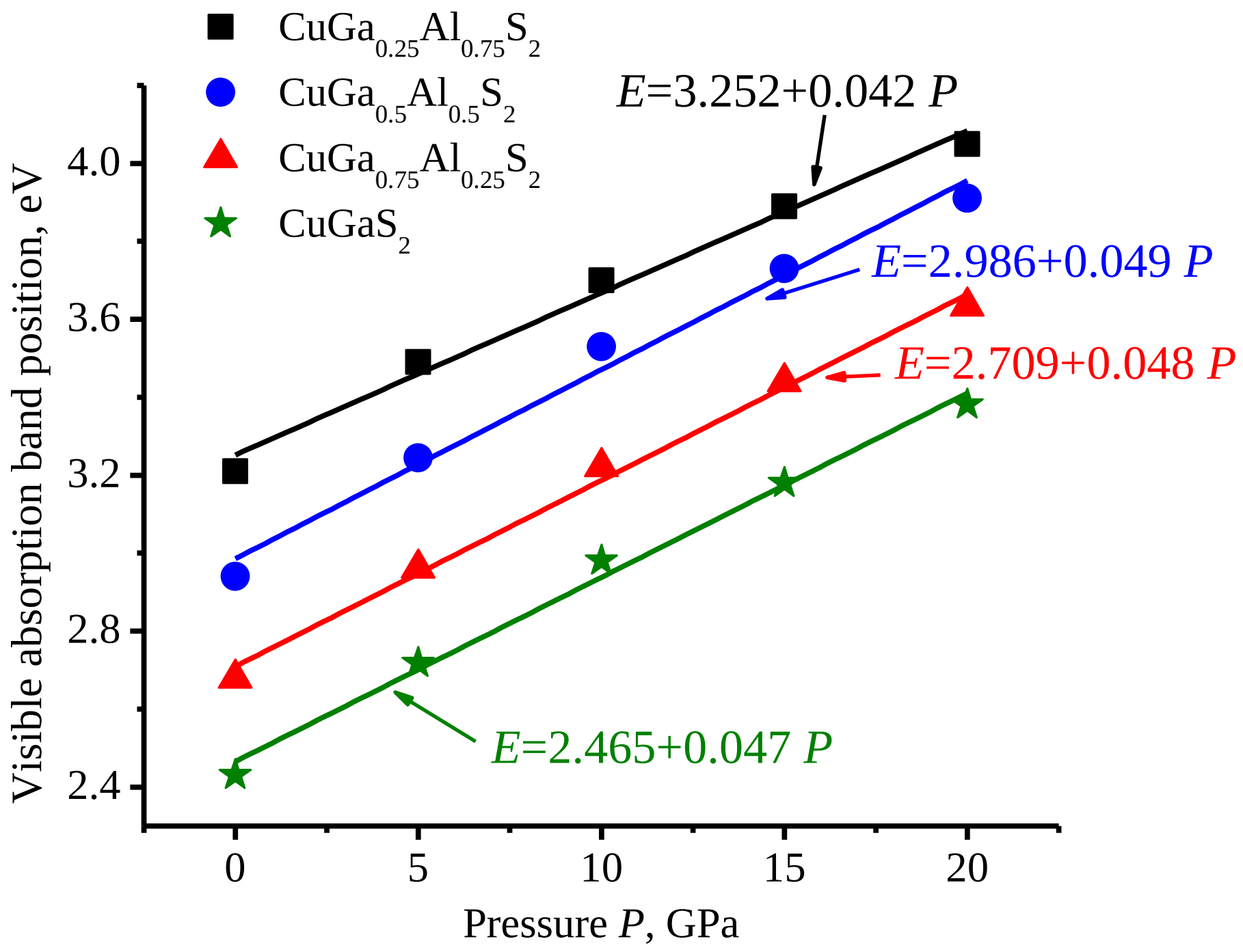

Fig. 10. Pressure dependence of the first absorption peak maximum for $CuGa_{1-x}Al_xS_2$ crystals.

Table 3. Mulliken charges and bond populations vs pressure in $CuGaS_2$

| Ions/bonds | 0 GPa | 5 GPa | 10 GPa | 15 GPa | 20 GPa |
|---|---|---|---|---|---|
| Cu | -0.10 | -0.12 | -0.14 | -0.16 | -0.18 |
| Ga | 0.82 | 0.81 | 0.82 | 0.82 | 0.82 |
| S | -0.36 | -0.35 | -0.35 | -0.33 | -0.32 |
| Cu-S | 0.48 | 0.50 | 0.51 | 0.51 | 0.52 |
| Ga-S | 0.41 | 0.42 | 0.43 | 0.43 | 0.44 |

Table 4. Mulliken charges and bond populations vs pressure in $CuAlS_2$

| Ions/bonds | 0 GPa | 5 GPa | 10 GPa | 15 GPa | 20 GPa |
|---|---|---|---|---|---|
| Cu | 0.19 | 0.18 | 0.17 | 0.16 | 0.15 |
| Al | 0.84 | 0.84 | 0.85 | 0.85 | 0.85 |
| S | -0.52 | -0.51 | -0.51 | -0.50 | -0.50 |
| Cu-S | 0.30 | 0.30 | 0.30 | 0.30 | 0.30 |
| Al-S | 0.58 | 0.59 | 0.60 | 0.61 | 0.61 |

As follows from Fig. 10, the first absorption peak for the $CuGa_{1-x}Al_xS_2$ ($x$=0, 0.25, 0.5, 0.75) crystals has almost the same pressure coefficient of about 0.042-0.049 eV/GPa. Application of an external pressure up to 10-15 GPa to pure $CuGaS_2$ sample would still keep it absorbing visible sunlight. Such a “critical” pressure determining visible light absorption would be decreased with increasing Al concentration.

Finally, Tables 3 and 4 show the calculated effective Mulliken charges [40] for pure $CuGaS_2$ and $CuAlS_2$, respectively, at different hydrostatic pressures. The most striking difference between these two compounds is a negative effective charge of Cu in $CuGaS_2$ and a positive effective charge of Cu in $CuAlS_2$, whereas effective charges of the Al and Ga ions are nearly equal in both materials. Difference in the charges of Cu ions in both pure hosts and requirement of electric neutrality result in a larger negative sulfur charge in $CuAlS_2$. The values of the Mulliken bond population are also collected in Tables 3-4. Although the quantitative analysis of the bond population is somewhat ambiguous, since the results depend on the choice of the basis set functions, the qualitative conclusions can be drawn. A high value of the bond population indicates a covalent bond, while a low value indicates an ionic interaction between the atoms or ions forming the bond in question [41]. Therefore, the Cu-S bonds are more covalent in $CuGaS_2$; the Ga-S bonds are less covalent than the Al-S bonds. It can be understood and explained by considering the electron configurations of these atoms: the $3d^{10}$ shell of Ga is completely filled and is lower in energy than the Cu $3d^{10}$ shell [10], whereas the Al 3s and 3p state are energetically closer to the S 3s, 3p states, favoring the overlap and covalent bond formation.

## 4. Conclusions

A detailed analysis of the composition and pressure influence on the structural, electronic and optical properties of $CuGa_{1-x}Al_xS_2$ chaplkopyrite semiconductor is reported in the present paper. The choice of the objects for the study is driven by the fact the neat $CuGaS_2$ thin films are used for solar cells applications. We have shown that substitution of 25 % of gallium by aluminum increases absorption of the visible light by about 6 %, which may lead to improved efficiency of the solar cells based on such a mixed compound.

The optimized lattice parameters for $CuGa_{1-x}Al_xS_2$ compounds were fitted to the linear functions of the aluminum composition *x*; the same was done for the band gaps and bulk moduli of all materials considered. This suggests the composition dependences on the structural, electronic and elastic properties of $CuGa_{1-x}Al_xS_2$ crystals in general obey the Vegard's law.

The optical properties (dielectric functions and absorption spectra) of all selected compounds were calculated. The emphasis was put on the spectral properties in the visible and near UV spectral ranges, since they are the most important for the solar cell applications. Dependence of the lowest in energy absorption peak (which falls with the visible spectral range until $x$~0.75) was shown to be perfectly described by the linear functions of the aluminum concentrations *x*. All dependencies of the band gap, bulk moduli, lattice parameters and position of the first absorption maxima give an opportunity of predicting all these properties for different aluminum concentrations or external pressures.

The results obtained in the present study may be useful for finding optimal conditions for applications of the $CuGa_{1-x}Al_xS_2$ thin films in photovoltaics.

## Acknowledgments

M. G. Brik thanks the support from European Social Fund's Doctoral Studies and Internationalisation Programme DoRa and the European Regional Development Fund (Centre of Excellence "Mesosystems: Theory and Applications", TK114). C.-G. Ma acknowledges the financial support by European Social Fund Grant No. GLOFY054MJD.

## References

[1] W.-J. Jeong, G.-C. Park, Solar Energy Materials & and Solar Cells 75 (2003) 93.
[2] A. Martí, D.F. Marrón, A. Luque, J. Appl. Phys. 103 (2008) 073706.